\documentclass[twocolumn,superscriptaddress,floatfix,preprintnumbers,prl ,nofootinbib,hyperref]{revtex4-2}
\usepackage[colorlinks=true,breaklinks=true]{hyperref}
\usepackage[utf8]{inputenc}
\hypersetup{allcolors=[rgb]{0.0 0.0 0.6},linkcolor=[rgb]{0.75 0.05 0.05}}
\usepackage{amsmath,amssymb}
\usepackage{mathtools}
\usepackage[normalem]{ulem}
\usepackage[dvipsnames]{xcolor}
\usepackage{float}
\usepackage{graphicx}
\usepackage{epsfig}
\usepackage{letltxmacro}
\LetLtxMacro{\oldcite}{\cite}
\renewcommand{\cite}[1]{\mbox{\oldcite{#1}}}
\usepackage{orcidlink}

\long\def\exclude#1{}

\newcommand{\rhoDM}{\rho_{_{\rm DM}}}
\newcommand{\mAp}{m_{A'}}

\DeclareMathOperator{\eV}{eV}

\DeclareSymbolFont{starfontsym}{OT1}{sts}{m}{n}
\DeclareMathSymbol{\mathTerra}{\mathord}{starfontsym}{76}

\newcommand{\beq}{\begin{equation}}
\newcommand{\eeq}{\end{equation}}

\def\ga{\,\,\raise0.14em\hbox{$>$}\kern-0.76em\lower0.28em\hbox
{$\sim$}\,\,}

\newcommand{\pd}{\partial}

\newcommand{\gayy}{g_{a\gamma \gamma}}

\newcommand{\wP}{\omega_{\rm pl}}

\long\def\exclude#1{}

\allowdisplaybreaks

\bibpunct{[}{]}{,}{n}{}{,}

\hypersetup{
    colorlinks=true,
    citecolor=[rgb]{.1, .7, .6},
    linkcolor=[rgb]{.2, .55, .95},
    filecolor=magenta,
    urlcolor=[rgb]{.1, .7, .6},
}

\begin{document}

\title{Resonant Conversion of Wave Dark Matter in the Ionosphere}

\author{Carl Beadle\,\orcidlink{0009-0004-9176-3827}} %\email{andrea.caputo@uv.es}
\affiliation{Departement de Physique Theorique, Universite de Geneve,
24 quai Ernest Ansermet, 1211 Geneve 4, Switzerland}

\author{Andrea Caputo\,\orcidlink{0000-0003-1122-6606}} %\email{andrea.caputo@uv.es}
\affiliation{Department of Theoretical Physics, CERN, Esplanade des Particules 1, P.O. Box 1211, Geneva 23, Switzerland}

\author{Sebastian A. R. Ellis\,\orcidlink{0000-0003-3611-2437}} %\email{andrea.caputo@uv.es}
\affiliation{Departement de Physique Theorique, Universite de Geneve,
24 quai Ernest Ansermet, 1211 Geneve 4, Switzerland}

%==========================

\begin{abstract}
We consider resonant wave-like dark matter conversion into low-frequency radio waves in the Earth's ionosphere. 
Resonant conversion occurs when the dark matter mass and the plasma frequency coincide, defining a range $m_{_{\rm DM}} \sim 10^{-9} - 10^{-8}\,\text{eV}$ where this approach is best suited.
Owing to the non-relativistic nature of dark matter and the typical variational scale of the Earth's ionosphere, the standard linearized approach to computing dark matter conversion is not suitable. We therefore solve a second-order boundary-value problem, effectively framing the ionosphere as a driven cavity filled with a positionally-varying plasma.
An electrically-small dipole antenna targeting the generated radio waves can be orders of magnitude more sensitive to dark photon and axion-like particle dark matter in the relevant mass range.The present study opens up a promising way of testing hitherto unexplored parameter space which could be further improved with a dedicated instrument.

\end{abstract}

\maketitle

\textit{Introduction.} The nature of dark matter (DM) remains a puzzle that requires an explanation from beyond the Standard Model (SM) of particle physics. Wave-like dark matter such as an axion-like particle (ALP) or a massive dark photon (DP) are well-motivated candidates~\cite{Nelson:2011sf,Arias:2012az,Graham:2015rva,Bastero-Gil:2018uel,Co:2018lka,Dror:2018pdh,Agrawal:2018vin,Bastero-Gil:2021wsf,Abbott:1982af,Preskill:1982cy}. Dark photons can naturally have a small coupling to SM photons through kinetic mixing~\cite{Holdom:1985ag}, while ALPs can have a CP-odd coupling to two SM photons~\cite{Dine:1982ah,Dine:1981rt,Kim:1979if,Shifman:1979if,Zhitnitsky:1980tq}. These two couplings are the subject of intensive theoretical and experimental work~\cite{Caputo:2024oqc, Safdi:2022xkm, OHare:2024nmr, DiLuzio:2020wdo, Irastorza:2018dyq}.

A massive DP could arise from an additional $U(1)$ gauge group broken by a compact scalar field, a possibility strongly motivated by UV completions of the SM~\cite{Dienes:1996zr,Giedt:2000bi,Gmeiner:2005vz,Dienes:2006ut,Dienes:2007ms,Arvanitaki:2009hb,Corti:2012kd,Taylor:2015ppa,Acharya:2016fge,Halverson:2018vbo,Acharya:2018deu}. The small kinetic mixing with the SM photon enables an
extensive experimental program to search for DP dark matter (see, e.g., ~\cite{Caputo:2021eaa} for a summary of ongoing efforts and experimental optimisation strategies). 
UV completions of the SM also often predict the existence of many ALPs~\cite{Svrcek:2006yi,Arvanitaki:2009fg,Demirtas:2021gsq,Broeckel:2021dpz}. These typically couple to photons, with a coupling strength that can be as large as $\gayy\sim 10^{-12}\,\text{GeV}^{-1}$~\cite{Halverson:2019cmy,Gendler:2023kjt}.

ALPs are CP-odd pseudoscalars while DPs are CP-even vectors, making these quite different dark matter candidates. However, they nevertheless often share similar phenomenology. We consider a possible signal due to resonant conversion of wave-like dark matter into radio waves in the Earth's ionosphere which is common to both ALPs and DPs. 
For the DP signal to exist, the presence of a plasma is sufficient, while for ALPs, a background magnetic field must also be present.
Both conditions are met in the weakly-ionised plasma of the Earth's ionosphere, where the Earth's small magnetic field ($B\sim 0.1\, \text{G}$) is present.

The structure of the interactions between DPs/ALPs and the SM photon are such that in a medium the mass eigenstates no longer correspond to the vacuum mass eigenstates. When the plasma frequency of the medium and the vacuum mass of the DM are degenerate, resonant level crossing between one state and the other can occur. For DPs, this condition has been exploited to study resonant conversion in various astrophysical enviromnents such as the solar corona~\cite{An:2020jmf,An:2023wij}, neutron star magnetospheres~\cite{Hardy:2022ufh}, or the intergalactic medium~\cite{Bolton:2022hpt, Caputo:2020rnx, Caputo:2020bdy, McDermott:2019lch}. For ALPs, this effect has also been studied in many astrophysical environments~\cite{Raffelt:1987im, Pshirkov:2007st, Hook:2018iia, Huang:2017egl, Witte:2021arp, Foster:2020pgt, Battye:2019aco, Leroy:2019ghm, battye2021robust}.

In this \textit{Letter} we propose searching for the conversion of dark matter in the Earth's own ionosphere. This approach has two advantageous properties: the ionosphere is well-studied and monitored (see \cite{Abdu2020-be} and references therein), allowing for a precise understanding of the conversion and propagation of the resulting radio waves; the peak plasma frequency in the ionosphere is $\omega_{\rm pl} \sim 10^{-8}\,\eV$, such that the mass range that can be probed is complementary to existing searches. 
Furthermore, galactic noise is reflected by the ionosphere, such that the dominant noise source is man-made or atmospheric, both of which can be monitored or even partially mitigated.
Several features of the ionosphere might allow for an improved ability to distinguish true signals from spurious ones.
For example, there is a daily modulation due to solar irradiation varying the free-electron number density, introducing a spectral feature in the true signal that would be absent for certain spurious signals.
Finally, for ALP searches, the dependence on the transverse component of the magnetic field makes the amplitude of the signal latitude-dependent.

\begin{figure*}[!t]
\centering
\includegraphics[width=0.49\textwidth]{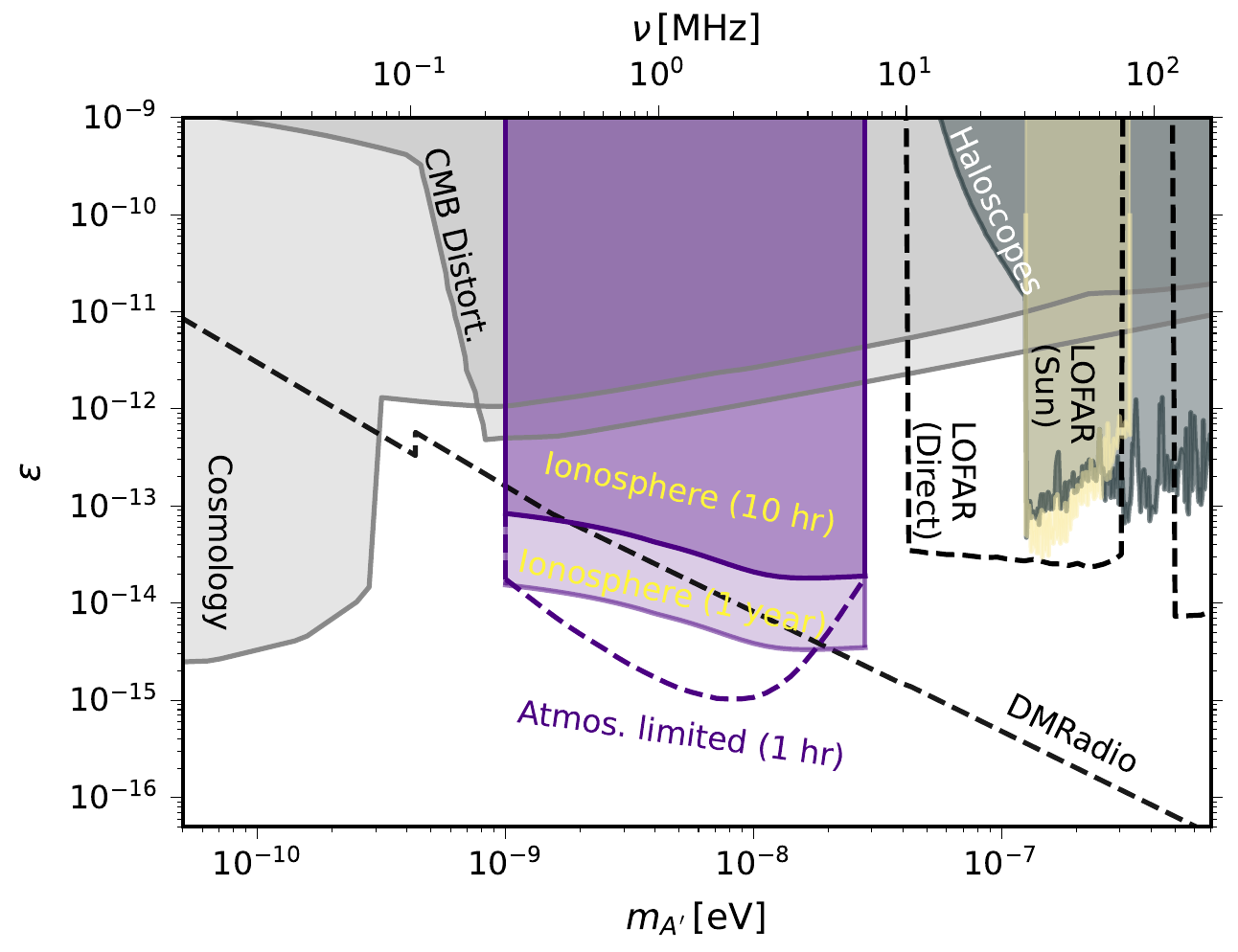}
\includegraphics[width=0.49\textwidth]{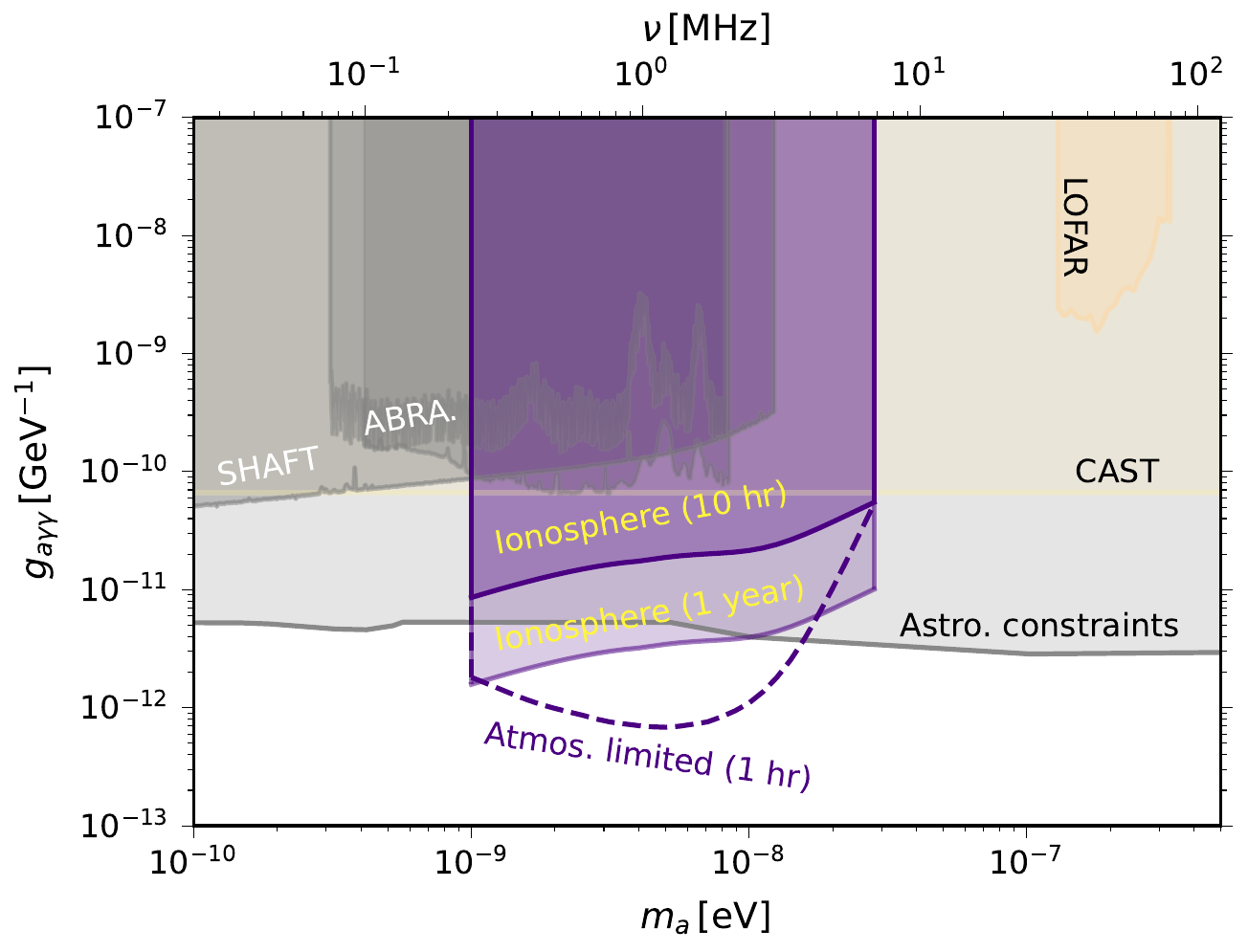}
\caption{(Left) Prospective reach in the DP kinetic mixing $\epsilon$ by considering a broadband search with integration time of $10$ hours and 1 year (solid purple curves). The dashed purple curve indicates the reach of 1 hour of observation when measurements are limited by atmospheric noise rather than man-made noise. The light grey region is excluded by cosmological probes~\cite{Caputo:2020bdy, McDermott:2019lch, Arias:2012az}, the dark grey region by Haloscopes, while the light gold region is excluded by LOFAR observation of the solar corona~\cite{An:2023wij}. The dashed black lines indicate possible future reach of LC-resonator DM-radio \cite{PhysRevD.92.075012}, as well as LOFAR reach for DP \textit{direct} detection in the antenna~\cite{An:2022hhb}. (Right) Projections for the axion to photon coupling $g_{a\gamma\gamma}$, with the same experimental set-up used for the DP. The light gray region is excluded by astrophysical probes~\cite{Noordhuis:2022ljw, Dessert:2022yqq, Davies:2022wvj, MAGIC:2024arq, Fermi-LAT:2016nkz}, the dark grey regions by terrestrial DM experiments ABRA\cite{Salemi_2021} and SHAFT\cite{Gramolin_2020}, while the light yellow region is excluded by CAST~\cite{CAST:2017uph}. The limit from LOFAR observation of the solar corona\cite{An:2023wij} are shown in light orange.}
\label{fig:Results}
\end{figure*}

\textit{DM conversion to electromagnetic waves.} The DP-photon system is described by the Lagrangian
\begin{equation}\label{Eq:Lagrangian}
\begin{split}
\mathcal{L} \supset &-\frac{1}{4}\left(F_{\mu \nu}F^{\mu \nu} - 2 \epsilon F'_{\mu \nu}F^{\mu \nu} + F'_{\mu \nu}F'^{\mu \nu} \right) \\
&+ \frac{1}{2}m_{A'}^2 A'_\mu A'^{\mu} - A_{\mu} \mathcal{J}^{\mu} \, ,
\end{split}
\end{equation}
where primed quantities are associated to the DP, while the axion Lagrangian is
\begin{equation}\label{Eq:LagrangianAxion}
\begin{split}
\mathcal{L} \supset & - \frac{1}{4} \Big(F_{\mu \nu}F^{\mu \nu} - 2 \, \partial_\mu a \, \partial^\mu a + g_{a\gamma\gamma}a F_{\mu \nu}\tilde{F}^{\mu \nu}\Big)   \\
& - \frac{1}{2}m_a^2 a^2 - A_{\mu} \mathcal{J}^{\mu} \, .
\end{split}
\end{equation}
The parameter $\epsilon$ is the kinetic mixing between the photon and the DP,  $g_{a\gamma\gamma}$ is the axion-photon coupling, while $m_{A'}$ and $m_a$ are the masses of DPs and axions, respectively. For convenience, we define the effective dark matter-photon coupling $g_{\rm eff} = \epsilon$ for DPs and $g_{\rm eff} = g_{a\gamma\gamma} |\mathbf{B}_T|/m_a$ for axions\footnote{In this letter we will take $|\mathbf{B}_T| \sim 0.4 \,\text{G}$ and assume it is homogeneous (a good approximation over the scales relevant to the ionosphere)~\cite{NOAA}.}. 

The evolution of the photon and dark matter system can be modelled as a two-state system of equations. While in vacuum the photon and dark matter are mass eigenstates, so no mixing can occur, in a medium such as a weakly-coupled plasma, the equations of motion of the two states become coupled through their interaction strength $g_{\rm eff}$. The form of the coupled equations implies that as long as $g_{\rm eff}$ is non-zero, resonant two-level crossing can occur when the effective photon mass (i.e. the plasma mass) and the dark matter mass are equivalent. If the spatial variations of the plasma frequency occur on scales much larger than the de Broglie wavelength of the DM, then the conversion probability is well approximated by the Landau-Zener formulae~\cite{Zener:1932ws, Landau:1932vnv, Parke:1986jy, Kuo:1989qe, An:2023wij}
\begin{align}\label{eq:ConversionProbSimple}
\begin{split}
    P_{\alpha\to\gamma} \simeq & (f_{\rm pol}\,\pi) \frac{ g_{\rm eff}^2\,m_\alpha}{v_r} \left\vert \frac{\pd \ln \wP^2}{\pd r}\right\vert^{-1}_{r_c} ,\
\end{split}
\end{align}
where $\alpha = A',~a$ depends on the dark matter candidate being considered. The polarisation fraction is $f_{\rm pol} = 2/3,~1$ for the DP and axion respectively.
The probability is evaluated at the conversion radius $r_c$ where $\wP(r_c) = m_{\alpha}$. The velocity factor $v_r \sim v_0$ is the radial component of the dark matter velocity, with $v_0 \simeq 220\,\text{km/s}$ the galactic dispersion velocity of dark matter. 

Unfortunately, for the Earth's ionosphere and for the dark matter masses of interest, 
the plasma frequency varies on a scale similar to or smaller than the de Broglie wavelength of the dark matter. As a result, the WKB approximation used in the derivation of the simplified formula in Eq.~\eqref{eq:ConversionProbSimple} does not hold, and the full second-order differential equations must be solved. We use the fact that the ionosphere plasma density has a strong gradient only along the $z$-direction to model the problem as a \textit{driven one-dimensional cavity filled with plasma}, where the driver is the DM field. The equation to be studied reduces to
\begin{equation}\label{eq:zmodeequation}
    \left[ \partial_z^2 + \omega^2 - \frac{\omega^2 \omega_{\rm pl}^2\left(z\right)}{\omega^2 + i \nu_c \omega} \,  \right] \mathbf{E}_T \left( z \right) = i \omega  \, g_{\rm eff} \, m_{\alpha}^2 \, \mathbf{V} \left( z \right),\\ 
\end{equation}
where $\mathbf{E}_T$ is the sourced electric field, $\mathbf{V} = \mathbf{A}^{\prime}_T~(a \, \mathbf{\hat{B}}_T)$ for the DP (axion), $\nu_c$ is the electron-ion collision frequency in the ionosphere, and $z$ is the height into the ionosphere as measured from the Earth's surface. 
The form of Eq.~\eqref{eq:zmodeequation} shows the salient aspects of the problem. When $\pd_z^2 + \omega^2 = m_\alpha^2 = \wP^2$, we see that there is a resonance as expected. Meanwhile, when $\wP^2\ll \omega^2$, we obtain the evolution of the transverse electric field as a function of $z$, subject to the appropriate boundary conditions. For the wavelengths of interest, the Earth acts as a good conductor, so that the field should vanish within one skin depth of the surface. Similarly, the plasma of the ionosphere behaves as a conductor for frequencies below $\wP$, imposing that the field should also vanish deep inside the plasma.

This 1D model breaks down if we consider DM waves with de Broglie wavelengths comparable to the Earth's radius, i.e. for $m_{\alpha} \lesssim 10^{-10}\,\text{eV}$. In practice, for DM masses below $m_\alpha \lesssim 10^{-9}\,\text{eV}$, our model of the ionosphere is a poor approximation of the real data~\cite{IRI16}, so we restrict ourselves to only considering masses above this value.
A technical description of 
our solution to Eq.~\eqref{eq:zmodeequation}
is provided in the Supplementary Material. 
Our formalism automatically takes into account all the wave propagation phenomena, including reflection, absorption and refraction of the electromagnetic (EM) waves that ultimately arrive at the detector. 

Fig.~\ref{fig:EM_energy} shows the EM energy density in natural units as a function of the ionosphere height for a fixed effective coupling, $g_{\rm eff} = 10^{-10}$. Different colours correspond to different DM masses; the solid curves are our numerical results, while the horizontal dashed lines show the result of applying Eq.~\eqref{eq:ConversionProbSimple}.
We notice that the resonant peak of each of our curves never deviates too much from the na\"ive calculation. However, the 
energy density near the Earth's surface, which is the quantity relevant for detection, is typically suppressed with respect to the peak. This is a particularly important effect for large masses, $\sim 10^{-8} \, \rm eV$, whose resonant conversion condition is only satisfied for the largest electron densities near the peak of the Chapman profile. An EM wave produced at that height undergoes many reflections as it propagates through the plasma, and its amplitude is therefore attenuated before it reaches the detector. The effect is less evident for smaller masses, where reflection plays only a minor role. 
The EM energy density near the Earth's surface is approximated to within $\sim10\,\%$ by the following sigmoid function
\begin{equation} \label{eq:sigmoid}
    \rho_{\rm EM} \simeq \frac{3 \times 10^{-23} \, \text{eV}^4 \left(\frac{g_{\rm eff}}{10^{-10}}\right)^2}{1 + \exp\Big[-\Big(\frac{m_{\alpha}}{2.3 \times 10^{-9} \text{eV}} - 3.8\Big)\Big]} \, ,
\end{equation}
which is valid for masses in the range $10^{-9}\,\leq m_\alpha/\text{eV} \lesssim 3\times 10^{-8}$. The lower boundary is defined by the aforementioned issues with the validity of our calculation, while the upper bound is defined by the peak values of the free electron number density. Ultimately, a detailed analysis taking into account the detector location and time could be performed using real ionosphere data~\cite{IRI16}, and could extend our sensitivity to smaller masses. We leave this to future work.

\begin{figure}
\begin{center}
\includegraphics[width=0.55\textwidth]{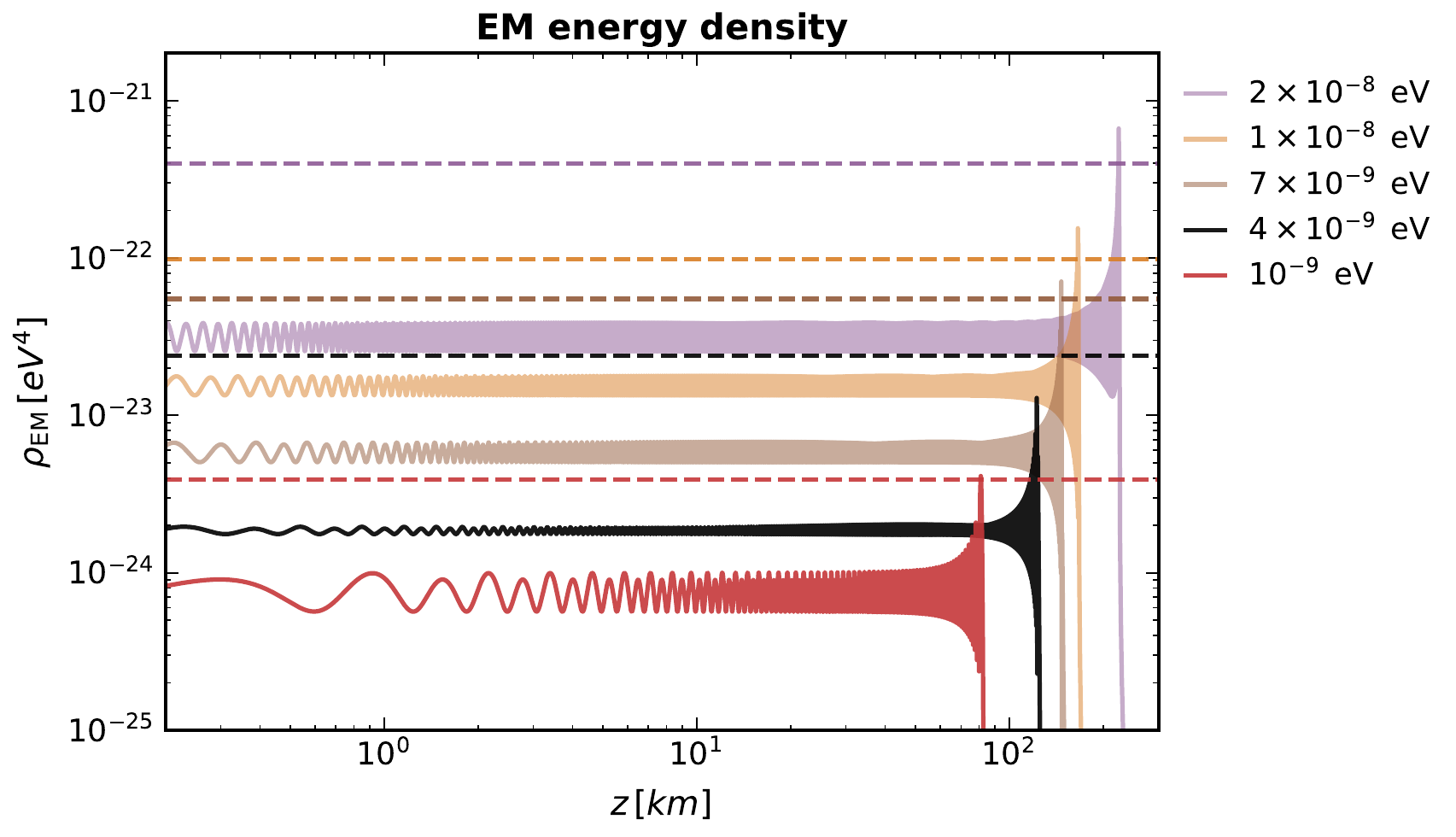}
\caption{EM energy density in natural units as a function of the distance z from the Earth surface. Different colors correspond to different DM masses, while the effective coupling is always fixed to $g_{\rm eff} = 10^{-10}$. The solid curves are our full numerical solutions, while the horizontal dashed lines correspond to the Landau-Zener conversion probability from Eq.~\eqref{eq:ConversionProbSimple}.}
\label{fig:EM_energy}
\end{center}
\end{figure}

\textit{Signal detection.} The EM radiation incident on the Earth's surface has a characteristic wavelength $\lambda \gg \text{m}$, and can therefore be detected with an electrically-small antenna~\cite{KMcD}. 
The signal approximated by Eq.~\eqref{eq:sigmoid} is the total integrated energy density. For detection, the more relevant quantity is the spectral density (SD) of the EM radiation $\mathcal{S}_{\rm sig}(\omega) \sim \rho_{\rm EM} f(\omega)$. The function $f(\omega)$ is approximately a Maxwell-Boltzmann distribution, normalised as $\int d\omega f(\omega) = 1$, which describes the frequency dispersion of the signal inherited from the dark matter velocity distribution. The signal is spread between frequencies $\omega \in m_\alpha[1, 1+\sigma^2/2]$, where $\sigma \sim 200\,\text{km/s}$ is the DM dispersion velocity. The bandwidth of the signal is thus narrow, and can be approximated as having an effective quality factor of $Q_{\rm sig} \sim 10^6$. Full details are given in the Supplementary Material.
 
The dominant noise at the relevant frequencies is from processes external to the receiver antenna. It is primarily a combination of atmospheric and man-made radiation. As a fiducial noise level, we adopt the man-made noise expected at a quiet rural location given by the International Telecommunication Union (ITU), see for example curve C of Fig.~2 of Ref.~\cite{ITU}. This can be characterised by the characteristic temperature of the Gaussian component of the noise
\begin{equation}\label{Eq:NoiseHumanT}
    T_{\rm N}(\nu) \simeq 6.1 \times 10^7 \left(\frac{\rm MHz}{\nu}\right)^{2.75} \, \text{K} \, .
\end{equation}
Under the assumption of an equivalent loss-free receiving antenna, this temperature can then be converted to a noise SD (see \textit{e.g} Ref.~\cite{KMcD} for a pedagogical derivation)
\begin{equation}\label{Eq:NoiseHuman}
    \mathcal{S}_{\rm N}(\nu) \simeq \frac{32}{3} \pi^2 \, \nu^2 \, T_{\rm N}(\nu) \, .
\end{equation}
A real device might contend not only with this typical man-made noise, but also with impulsive components at particular frequencies. Furthermore, atmospheric noise leads to a temperature that can vary significantly depending on weather conditions, sometimes exceeding typical man-made noise by many orders of magnitude~\cite{ITU}. 

Both the signal and the noise are external to the antenna, and are filtered by the same transfer function determining the antenna response, which therefore does not enter the SNR. As a result, the optimal signal-to-noise ratio (SNR) is given by~\cite{Chaudhuri:2018rqn, Foster:2017hbq}
\begin{equation}
\label{eq:SNR}
\text{SNR} = \left[ t_{\text{int}} \int_{0}^{\infty} d\nu \left(\frac{\mathcal{S}_{\text{Sig}}}{\mathcal{S}_{\text{N}}}\right)^2 \right]^{1/2},
\end{equation}
where $t_{\rm int}$ is the integration time of our measurement (assumed to be larger than the dark matter coherence time). If the receiver antenna is critically coupled, it will have a narrow bandwidth owing to the small radiation resistance. As a result, it is optimal to couple the antenna to an additional in-series resistance. In the Supplementary Material we provide a simple model for an RLC circuit that allows to broaden the frequency response up to $\Delta\nu \sim \text{MHz}$. The circuit we describe, and the value of its parameters, are similar to those of very old radio missions~\cite{Cane79, Novaco1978}. The result of this broad frequency response is that in order to scan an e-fold in DM mass $t_e$, an integration time at a given frequency of $t_{\rm int} \sim t_e \,\mathrm{min} \left( 1, 2\, \pi\, \Delta\nu / m_\alpha \right)$ is required.

Fig.~\ref{fig:Results} shows our fiducial prospects (solid purple lines) for a broadband search with 1 MHz bandwidth, for $10$ hours and one year of e-fold time, for both DPs (left panel) and axions (right panel). In both panels light grey regions are excluded by cosmological and astrophysical probes~\cite{Caputo:2020bdy, McDermott:2019lch, Arias:2012az, Noordhuis:2022ljw, Dessert:2022yqq, Davies:2022wvj, MAGIC:2024arq, Fermi-LAT:2016nkz}. Observations by LOFAR of the solar corona are shown in light orange~\cite{An:2023wij} in both panels. For the DP panel the dark grey region is excluded by Haloscopes. The dashed black lines indicate possible future sensitivity of DM Radio~\cite{PhysRevD.92.075012}, as well as LOFAR sensitivity to \textit{direct} absorption by the antenna. For the axion panel, the dark grey regions are excluded by terrestrial DM experiments ABRA~\cite{Salemi_2021} and SHAFT~\cite{Gramolin_2020}, while the light yellow region is excluded by CAST~\cite{CAST:2017uph}. 

In case man-made noise can be mitigated, we also show a dashed purple curve corresponding to the typical atmospheric noise in Western Australia around midday on a winter day (see Fig. 18 of Ref.~\cite{ITU}), assuming \textit{a single hour} of e-fold time.

\textit{Conclusion.} In this work we proposed a new way to detect bosonic dark matter with mass $m_{\alpha} \lesssim 3 \times 10^{-8}$ eV, \textit{i.e} below the typical maximum ionosphere plasma frequency. When DM waves pass through the ionosphere of the Earth, they can get resonantly converted into radio waves that are detectable by a small meter-scale antenna. Our projections suggest many decades of DP parameter space could be probed in just a few hours of observation time. The small magnetic field of the Earth affects the sensitivity to axions, but we nevertheless project that a similar setup can improve on the best laboratory constraints, and possibly the best astrophysical constraints. 

The present work naturally leaves open questions to be addressed in future studies. Fully characterising the electrical and physical properties of the antenna should be done. The location of the antenna can also be optimised, depending on man-made and atmospheric noise, as well as the Earth's magnetic field for the axion.
%Then, w
With a precise detector design and location in mind, a more realistic modelling of the ionosphere plasma frequency using available data~\cite{IRI16} can be performed, accounting for diurnal variations. The diurnal variation can be used to look for modulations of our signal, 
which could be useful in discriminating it from backgrounds. Moreover, our signal can be characterised by the 
propagation of the signal radially towards the Earth's surface, $\mathbf{k} \propto \hat{\mathbf{r}}$, imprinted by the large plasma gradient in this direction. 

Finally, given the simplicity, (small) size and low cost of the proposed antennae, we envision the use of an array of antennae operating in an interferometric mode. Placing $N$ antennae $\sim \mathcal{O}(10) \, \rm km$ from each other can improve the signal to noise ratio by at least a factor $\sqrt{N}$. The coherence length of the DM signal would exceed the antenna separation, while man-made noise varies more over these scales, thus the potential for improvement is greater if it enables the subtraction of man-made noise sources.

\textit{Acknowledgments.} We thank Itay Bloch, Sam Witte for multiple enlightening conversations, and Nick Rodd and Kevin Zhou for comments on the manuscript. We particularly thank Pietro Bolli, Giulia Macario, Jader Monari, Federico Perini for illuminating discussions about low-frequency antennas and radio background noise. The work of CB and SARE was supported by SNF Ambizione grant PZ00P2\_193322, \textit{New frontiers from sub-eV to super-TeV}.

%%%%%%%%%%%%%%%%%%
%%%%%%%%%%%%%%%%%%

\bibliographystyle{bibi}
\bibliography{biblio}

\appendix

\clearpage
\newpage
\maketitle
\onecolumngrid
\begin{center}
\textbf{\large Resonant Conversion of Wave Dark Matter in the Ionosphere} \\ 
\vspace{0.05in}
{ \it \large Supplemental Material}\\ 
\vspace{0.05in}
{}
{Carl Beadle, Andrea Caputo, and Sebastian A. R. Ellis}

\end{center}
%%%%%%%%%% Merge with supplemental materials %%%%%%%%%%
\setcounter{equation}{0}
\setcounter{figure}{0}
\setcounter{table}{0}
\setcounter{section}{1}
\renewcommand{\theequation}{S\arabic{equation}}
\renewcommand{\thefigure}{S\arabic{figure}}
\renewcommand{\thetable}{S\arabic{table}}
%\interfootnotelinepenalty=100000 % force footnotes to be together

%\setstretch{1.1}

Below we provide additional details of the calculations leading to the results presented in the main text. In particular, we describe the ionosphere plasma model we adopt and describe how we compute the electromagnetic field amplitude at the Earth's surface. 
We also explain how we model the frequency spread of the signal, and compare the signal from resonant conversion in the ionosphere with that seen directly by an antenna if the conversion was neglected. Finally, we describe a simple antenna circuit that can achieve the broadband sensitivity we argue for in the main text.

\vspace{0.5cm}

\section{Ionosphere modelling: the Chapman profile}

\begin{figure}[h]
    \centering
    \includegraphics[width = 0.45\textwidth]{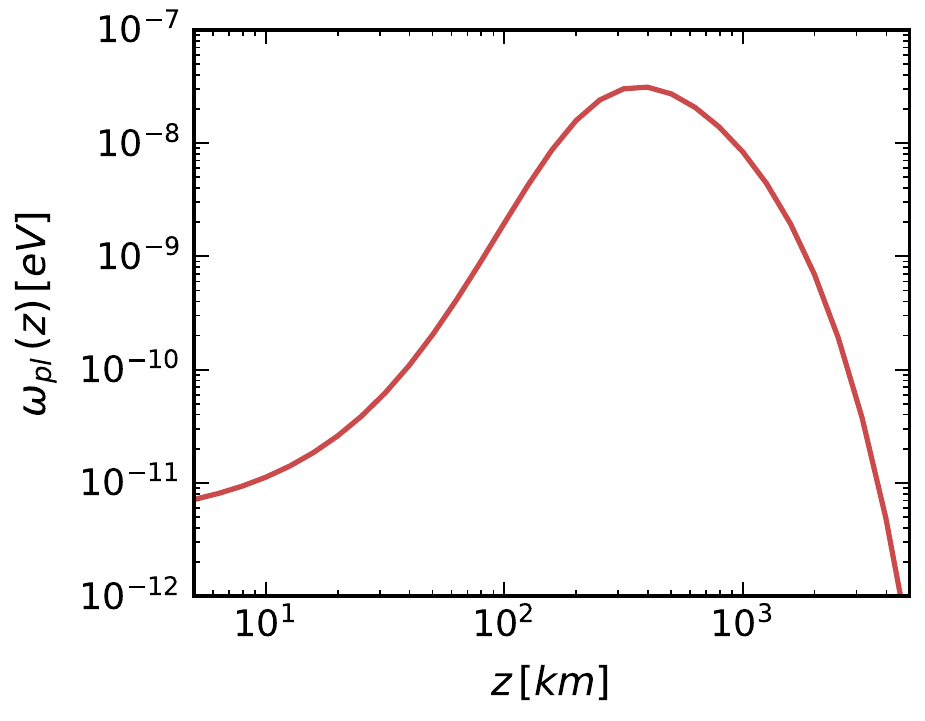}
    \caption{Simple Chapman profile for the plasma frequency in the ionosphere as a function of the distance from the Earth surface.}
    \label{fig:Chapman}
\end{figure}

A very simple parameterisation of the ionosphere electron density is the so-called Chapman model~\cite{Chapman_1931}. This is the model we use to obtain our results in the main text. The model has three parameters that must be provided as input: the maximal free electron density $n_{\rm max}$, the scale height $H$ and the maximal height $z_{\rm max}$. 
In terms of these three parameters, the free electron number density as a function of height can be expressed as~\cite{Chapman2003}
\begin{equation}
n_e(z) = n_{\rm max} \exp \left[\frac{1}{2}\left(1-\frac{z-z_{\rm max}}{H}\right) - \exp\left(-\frac{z - z_{\rm max}}{H}\right)\right],
\end{equation}
where $z$ is the distance from the Earth surface, and we have set $n_{\rm max} = 10^6 \, \text{cm}^{-3}$, $H = 100 \, \text{km}$ and $z_{\rm max} = 300 \, \text{km}$. In Fig.~\ref{fig:Chapman} we plot the corresponding plasma frequency, $\omega_{\rm pl}(z) = \sqrt{ 4 \pi \, n_e(z) \alpha/m_e}$, where $m_e$ is the electron mass and $\alpha$ the electromagnetic structure constant.
\noindent As discussed in the main text, this model is a reasonable approximation to the daytime free electron number density for plasma frequencies above $\omega_{\rm pl}(z)\gtrsim 10^{-9}\,\text{eV}$ only.

\section{Method for solving for electromagnetic field and energy density at detector}

In this section, we describe the method we use to solve for the electromagnetic fields and energy density that arrives at the detector.

For the mass range under consideration, the  
phase-space density of the dark matter particles
is high, 
such that it is appropriate to treat the system classically. 
We consider the dark matter field to constitute the full local dark matter abundance, such that we may write it as follows

\begin{equation}\label{eq:planewaveDP}
    \left\{ \begin{array}{c}
        \mathbf{A}^{\prime} \left( \mathbf{x}, t \right) \\
        a \left( \mathbf{x}, t \right)
    \end{array} \right\}
    = \frac{\sqrt{\rhoDM}}{m_\alpha} \sum_l f_l \left\{ \begin{array}{c}
        \mathbf{n}_l \\
        1
    \end{array} \right\} \exp \left( - i \omega_l t + i \mathbf{k}_l \cdot \mathbf{x} + i \phi_l \right),
\end{equation}
where $m_\alpha$ is the mass of $\alpha = A',~a$ where appropriate, and $f_l$ is the combination of the velocity distribution in a local frame and a random variable drawn from a Rayleigh distribution. The quantity $\phi_l$ is a random phase, while $\mathbf{n}_l$ specifies the polarisation of the vector field in the case of the DP.
The dispersion of the dark matter wave oscillation frequency $\omega_l$ (discussed in greater detail below) is such that there is a natural coherence time $\tau_c \sim 2\pi/ m_\alpha v^2$ where $v\sim 200\,\text{km/s}$. For the masses of interest, this coherence time is about one second. The expected data-taking campaign will involve recording data for much longer timescales, such that we can approximate the DM wave as having a fixed amplitude $\sqrt{\rhoDM}/m_\alpha$.

We can treat the presence of wave-like DM as leading to an additional \emph{effective} current density $j^\mu_{\rm eff} = (\rho_{\rm eff},~\mathbf{j}_{\rm eff})$
in Maxwell's equations
\begin{alignat*}{2}\label{eq:ModMax}
	\nabla \cdot \mathbf{D} &= \rho - \rho_{\rm eff},\;\;  &&\nabla \times \mathbf{H} - \partial_t \mathbf{D} = \mathbf{j} - \mathbf{j}_{\rm eff}, \\
	\nabla \cdot \mathbf{B} &= 0,  &&\nabla \times \mathbf{E} + \partial_t \mathbf{B} = 0 \ .
\end{alignat*}
The effective charge density is $\rho_{\rm eff} = \epsilon \, \mAp^2 A^{\prime \, 0}$ for the DP and $\rho_{\rm eff} = g_{a\gamma\gamma} \mathbf{B} \cdot \nabla a $ for the axion. The effective 3-current density can be written in terms of the effective coupling and a vector $\mathbf{V}$ defined in the main text, $\mathbf{j}_{\rm eff} = \, g_{\rm eff} \, m_{\alpha}^2 \mathbf{V}$. We remind the reader that $\mathbf{V} = \mathbf{A}^{\prime}_T$ for the DP, and $\mathbf{V} = a \, \mathbf{\hat{B}}_T$ for the axion. Furthermore, we recall the definition $g_{\rm eff} = \epsilon$ for DPs and $g_{\rm eff} = g_{a\gamma\gamma} |\mathbf{B}_T|/m_a$ for axions.

The Drude model allows for a simplified characterisation of the motion of electrons in the plasma in the presence of EM fields. The model combines the Lorentz force law with a collision term that acounts for electron-ion collisions. The model allows us to solve in frequency space for the motion of the electrons, and thereby derive the physical EM current $\mathbf{j}$ entering in Maxwell's equations. Neglecting the Earth's magnetic field,\footnote{This is not always a good approximation in the parameter space we consider. The main effect of including the magnetic field would be that the resulting EM radiation gets polarised. However, without detailed knowledge of the antenna type and its placement, any inclusion of magnetic field effects on the resulting EM radiation is premature.} we can solve for the average momentum of electrons
\begin{align}
    \langle \mathbf{p} \rangle \simeq - \frac{i}{\omega + i \nu_c} \frac{e}{m_e}\mathbf{E} \ .
\end{align}
This can then be used to find the average current density
\begin{equation}
    \mathbf{J} = - e\langle \mathbf{p} \rangle \frac{n_e}{m_e} \,  = %\begin{cases}
    % \frac{i \, \omega_p^2 }{\omega \pm \omega_b + i \nu} \, E_{\pm}\\[10pt]
    % \frac{i \, \omega_p^2}{\omega + i \nu} \, E_z\end{cases}. 
    \frac{i \omega_{\rm pl}^2}{\omega + i \nu_c} \mathbf{E} \ ,
\end{equation}
where we have used the definition of the plasma frequency in the second equality.
The current density can be related to the charge density by the continuity equation in frequency space.
Assuming the entirety of the effect of charged or polarised matter is contained in $\mathbf{j},~\rho$, we can further simplify Maxwell's equations by setting $\mathbf{H} = \mathbf{B}$ and $\mathbf{D} = \mathbf{E}$. This approach is equivalent to setting up the problem with both a conductivity and a polarization tensor and assuming no free charges or currents are present.
Unless otherwise stated, from here on, we are working with the frequency modes of any fields.

One can keep track of the relevant degrees of freedom in the problem by performing a Helmholtz decomposition,
\begin{equation}\label{eq:Helmholtz}
	\mathbf{E} = \mathbf{E}_T + \mathbf{E}_L \;\; \text{such that} \;\; \nabla \cdot \mathbf{E}_T = 0 \;\; \text{and} \;\; \nabla \times \mathbf{E}_L = \mathbf{0} \ .
\end{equation}

We will only consider transverse modes, as these are the ones relevant to detection on Earth. The second order differential equation governing their evolution is obtained by combining Amp\'ere's and Faraday's laws, yielding
\begin{equation}\label{eq:transverseequation}
	\nabla^2 \, \mathbf{E}_T + \omega^2 \left( 1 - \frac{\omega_p^2}{\omega^2 + i \, \nu \, \omega } \right) \, \mathbf{E}_T = i \, g_{\rm eff} \, m_{\rm DM}^2 \, \omega \, \mathbf{V}, \\
\end{equation}
which is the 3D generalisation of Eq.~\eqref{eq:zmodeequation} in the main text.
In order to make further progress some information about the plasma needs to be specified. 

A natural approximation to make in the case of the ionosphere is that the number density of electrons is primarily a function of height from the surface of the Earth. For the 
masses under consideration, we are justified in ignoring effects coming from the curvature of the Earth; they will contribute at most $\mathcal{O}(1/ m_{\rm DM} \, v \, R_{\oplus})$ corrections. The combination of these assumptions leads us to work in a 1D approximation where the only relevant variation is in the $z$-direction, defined as the height above the surface of the Earth. 
One can now see that there is a good translational symmetry in the transverse directions, implying conservation of momentum in these directions.
An immediate consequence of this will be the significant refraction of light rays towards the $z$-direction as they propagate downwards.
A photon near the Earth's surface should approximately satisfy the dispersion relation: $\omega^2 \approx \mathbf{k}^2$.
Given that the photon produced as a result of the DM effective current should inherit both the frequency \textit{and} transverse momentum of the DM, the only way this relation can be satisfied is if the momentum in the $z$-direction is of order $m_{\alpha}$, which is $\sim 10^3$ times larger than the value in the transverse directions.
An alternative, but equivalent, way to see the same effect is by considering the relation between the plasma frequency and the refractive index of the ionosphere and using Snell's law.

It is now very natural to decompose the fields of Eq.~\eqref{eq:transverseequation} into plane waves in the transverse directions and some generic function of the $z$-direction, resulting in: 
\begin{equation}\label{eq:zmodeequation2}
    \left[ \partial_z^2 + \omega^2 -\frac{\omega^2}{\omega^2 + i \nu \omega} \, \omega_p^2 \left(z\right) \right] \, \mathbf{E}_T \left( z \right) = i \, g_{\rm eff} \, m_{\rm DM}^2 \, \omega \, \mathbf{V}\left( z \right).\\ 
\end{equation}
The problem has now been reduced to finding the form of these modes.
For a totally generic plasma frequency profile this is not a trivial problem, as the gradient and any possible turning points of the plasma will influence the amplitude of a wave arriving at the detector.
For the radio wavelengths of interest, the surface of Earth is effectively a perfect conductor, thereby imposing an important boundary condition on the problem.

In order to solve Eq.~\eqref{eq:zmodeequation2} we adopt a finite difference method. We first discretize the equation as
\begin{equation}
\frac{E_{i-1} -2 E_i + E_{i+1}}{\Delta z^2} + E_i \Big(\omega^2 - \frac{\omega^2 \omega_{\rm pl, i}^2}{\omega^2 + i \nu_i \omega}\Big) = i \, g_{\rm eff} \, \omega \, m_{\rm DM}^2 V_i,
\end{equation}
where the subscript $i$ indicates the $i$th position on a grid along the z-axis. We are then left with an algebraic system of the type
\begin{equation}
A_i \, E_{i-1} + B_i \, E_i + C_i \, E_{i+1} - D_i = 0,
\end{equation}
where $A_i = C_i = 1/\Delta z^2$, $B_i = -\frac{2}{\Delta z^2} + \Big(\omega^2 - \frac{\omega^2}{\omega^2 + i \nu_i \omega}\omega_{\rm pl, i}^2\Big)$ and $D_i = i \, g_{\rm eff} \, \omega \, m_{\rm DM}^2 V_i$. We thus have to solve a tridiagonal system of equations, and to do so we employ the well-known \href{https://en.wikipedia.org/wiki/Tridiagonal_matrix_algorithm}{Thomas algorithm}.

\section{Frequency spread of the DM}

In this section we give some details about the DM energy distribution which enters the signal PSD in the main text. We make the simplifying approximation that the DM is well-described by a gas of non-relativistic particles and is thus characterised by the following momentum distribution~\cite{Krauss:1985ub}:
\begin{equation}\label{eq:MomentumMB}
    f \left( \mathbf{p} \right) = \frac{1}{\left(  \frac{2 \pi }{3} m^2 \langle v^2 \rangle\right)^{3/2}} \, \exp \left[ - \frac{\mathbf{p}^2}{\frac{2}{3} m \langle v^2 \rangle} \right],
\end{equation}
where we have fixed the normalisation such that the distribution integrates to one and the rms momentum is given by $\langle \mathbf{p}^2 \rangle = m^2 \langle v^2 \rangle$.
Here $\langle v^2 \rangle$ denotes the mean-square velocity, which in the case of isotropic distribution is related to the velocity dispersion by $\langle v^2 \rangle = 3 \, \sigma^2$, where $\sigma = \sqrt{3/2} \, v_c$, with $v_c = 220 \, \rm km/s$ being the circular velocity for the Milky Way~\cite{Chaudhuri:2018rqn, Herzog-Arbeitman:2017zbm}.
We then make the identification $f(\omega) \, \mathrm{d}\omega = f( \mathbf{p} ) \, \mathrm{d}^3 p$ to find the distribution in frequency space, namely:
\begin{equation}\label{eq:FreqMB}
    f \left( \omega \right) = \frac{1}{\sqrt{\pi}} \, \frac{\sqrt{\left| \omega \right| - m}}{\left(  \frac{1 }{3} m^2 \langle v^2 \rangle\right)^{3/2}}  \, \exp \left[ - \frac{\left| \omega \right| - m}{\frac{1 }{3} m^2 \langle v^2 \rangle} \right] \, \theta \left( \left| \omega \right| - m \right).
\end{equation}
We see that this is the same as previously found in the literature \cite{PhysRevLett.55.1797}.
As mentioned in the main text, the signal should inherit this frequency spread.
The PSD of the signal E-field may then be constructed as:
\begin{equation}
    \mathcal{S}_{\mathrm{sig}} \left( \omega \right) = 4 \pi \, \left( \frac{\omega}{m} \right)^2 \, f \left( \omega \right) \rho_{\mathrm{sig}} , \\
\end{equation}
complying with our usual conventions for PSDs.
A straightforward calculation shows that this PSD is peaked at $\omega \approx m \left( 1 + \frac{1}{6} \langle v^2 \rangle \right)$; assuming that the signal energy density varies more slowly in $\omega$ than $f\left(\omega\right)$.
From here we may find the bandwidth of this PSD by the usual FWHM criterion, where we see that it given by the expected: $\Delta \omega \approx m \langle v^2 \rangle$.

\section{Comparison with direct detection}

The same antenna used to detect radiation resulting from the resonant conversion of wave-like DM in the ionosphere can also be sensitive to the non-resonantly-converted signal that is anyway present on Earth.

In the case of dark photon DM, this is especially straightforward to understand: if there is a non-zero $A'$ amplitude in the vicinity of the antenna, it can couple to charges in the antenna and generate a current. The signal PSD associated to this effect is
\begin{align}
    \mathcal{S}_{\rm sig,\, DD}(\omega) \simeq \epsilon^2 \rhoDM f(\omega) \ ,
\end{align}
and would enter Eq.~\eqref{eq:SNR} as an additional contribution to the signal PSD in the numerator. We notice that this PSD and the resonant conversion signal PSD share the same spread in frequency space, $f(\omega)$, such that the comparison between the two signals amounts to a comparison between $\rhoDM$ and Eq.~\eqref{eq:sigmoid}. For $g_{\rm eff} = \epsilon = 10^{-10}$, we find that $\epsilon^2 \rhoDM \sim 3 \times 10^{-26}\,\text{eV}^4$, and is therefore between 2 and 3 orders of magnitude smaller than the EM energy density due to resonant conversion given in Eq.~\eqref{eq:sigmoid} for the mass range being considered.

For axion DM, the analogous effect can be estimated by arguing that axions can convert into an electric field parallel to the Earth's magnetic field of a magnitude $E_a \sim \gayy a B$, such that the direct detection signal PSD is approximately
\begin{align}
    \mathcal{S}_{\rm sig,\, DD}(\omega) \simeq \gayy^2 B_\oplus^2 \, \frac{\rhoDM}{m_a^2}f(\omega) \ .
\end{align}
The prefactor without the frequency spread function can be compared with Eq.~\eqref{eq:sigmoid}. This is the same comparison as for the dark photon above, indicating that once again, the resonant conversion EM signal PSD is between 2 and 3 orders of magnitude larger than the direct absorption PSD.

\section{Modelling of transfer function in a simple antenna circuit}

Here we provide a simple description of a small linear antenna that could be used for detection of the DM signal. This is not intended to be exhaustive of the list of possibilities to detect our signal, but it gives an idea of the simplicity of a possible detector.

The antenna and read-out circuit combined are treated as a simple, in-series RLC-circuit, where the driving voltage is provided by the antenna.
There is a characteristic impedance $Z_A = R_A + X_A$ for the antenna element, determined primarily by its geometry and material properties.
We consider a linear antenna of length $h$, cross-sectional area $A$, resistance $R_A \equiv R_{\rm Ohm} + R_{\rm rad}$ and reactance $X_A (\lambda)$.
In this case, we are working in the ``electrically small" limit for the antenna meaning that the radiation resistance reads:
\begin{equation}
    R_{\rm rad} = \frac{2 \pi \, Z_0}{3} \left(\frac{h}{\lambda}\right)^2 \simeq 0.05 \, \Omega \, \left(\frac{h}{\rm m}\right)^2 \left(\frac{m_{\alpha}}{10^{-8} \rm eV}\right)^2,
\end{equation}
where we identified $\lambda = 2 \pi/m_{\alpha}$ and the reactance of the antenna is largely dominated by its effective capacitance:
\begin{equation}
    X_A(\lambda) \approx - \frac{Z_0}{2 \pi } \frac{\lambda}{C_A} \approx - \frac{Z_0}{\pi^2}\left(\frac{\lambda}{2 \, h}\right) \ln\left(\frac{h}{ \sqrt{A/\pi}}\right).
\end{equation}
The remaining circuit elements are well-modelled by a resistive load element `$R_L$' and an inductive element `$L$'.

The power dissipated across the load may be computed by finding the voltage across it using Kirchhoff's laws and the standard relation for power in a circuit element.
We see that the current flowing through any given element is:
\begin{equation}\label{apeq:current}
   I\left( \omega \right) = \frac{i \, \omega}{L \left( \omega^2 - \omega_0^2 + i \, \omega \, \Delta\nu \right)} \, V_d \left( \omega \right), \\
\end{equation}
where the circuit resonant frequency is $\omega_0^2 \equiv 1 / \left( C_A L \right)$ and the bandwidth is $\Delta\nu \equiv \left( R_A + R_L \right) / L$. The result is that the power dissipated is
\begin{equation}
    P_L = \int \mathrm{d} \omega \, \frac{\omega^2 \, h^2}{R_L \, L^2 \, \left[ \left( \omega^2 - \omega_0^2 \right)^2 + \omega^2 \, \Delta\nu^2 \right]} S_E \left( \omega \right), \\
\end{equation}
where $S_E$ is the PSD of the electric field incident on the antenna, consisting of the signal and any external noise sources.
We note that the bandwidth is dominated by a constant $R_L > R_A$ for resistive elements easily available for purchase by amateurs, leading to a broadband detector response.
As discussed in the main text, by far the dominant noise source is extrinsic to the detector, and therefore is filtered by the same detector response as the signal. 
This transfer function of the antenna, i.e. the prefactor of $S_E(\omega)$ in the integrand above, therefore factorises in the integrand of the SNR, Eq.~\eqref{eq:SNR} in the main text. However, the bandwidth of the receiver $\Delta \nu$ affects the scan rate, which enters the expression for the integration time $t_{\rm int}$ in Eq.~\eqref{eq:SNR}. 
\begin{figure}[t]
    \centering
    \includegraphics[width=0.6\textwidth]{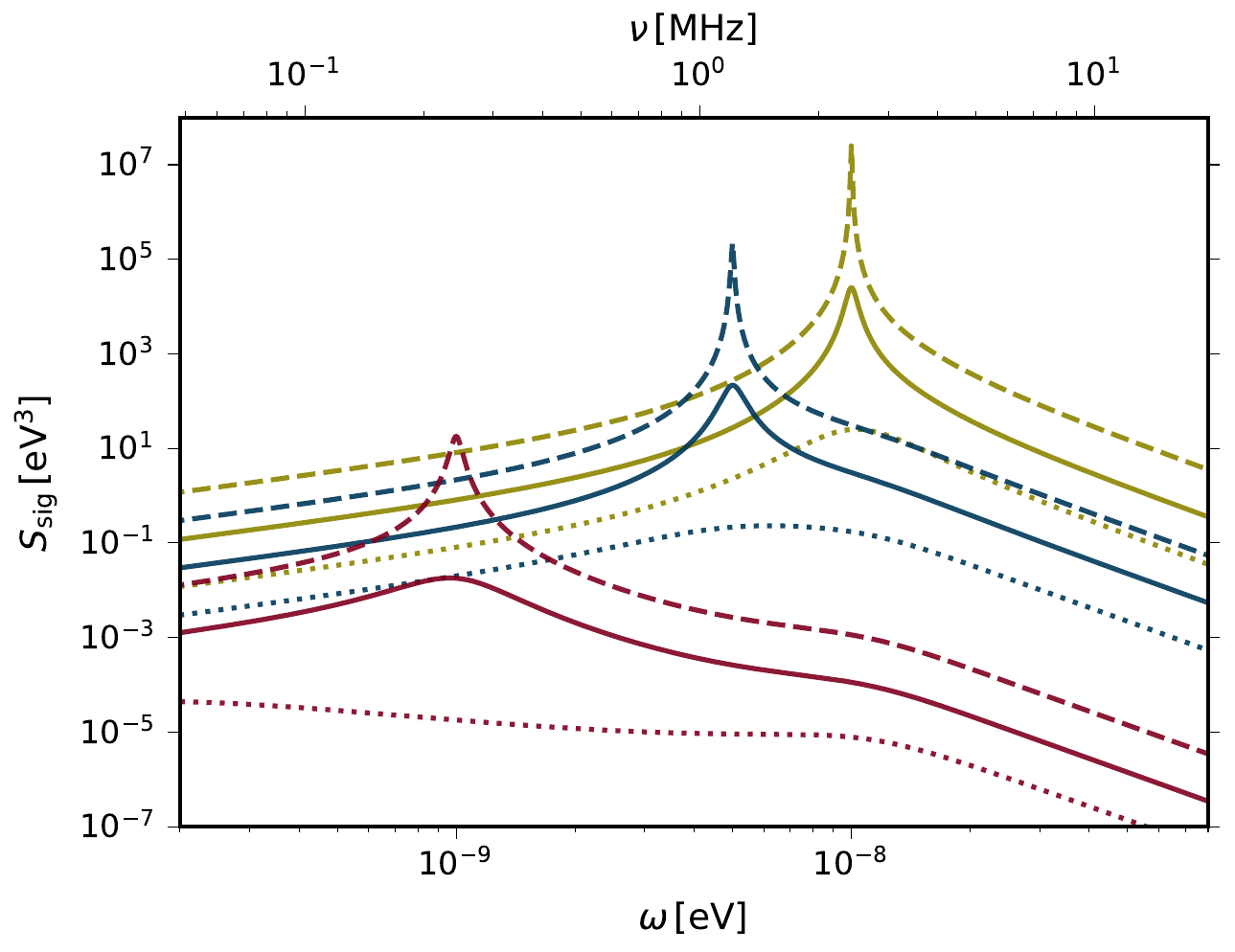}
    \caption{The value of the signal PSD fed through the transfer function of the antenna circuit assuming the signal is peaked at frequency $\omega$, for all curves we take $g_{\rm eff} = 10^{-10}$. The value of $\omega_0$ is $1\cdot 10^{-9} \mathrm{eV}$ for red curves, $5\cdot 10^{-9} \mathrm{eV}$ for blue and $1\cdot 10^{-8} \mathrm{eV}$ for golden curves. The bandwidth of the circuit is $10 \mathrm{MHz}$ for dashed lines, $1 \mathrm{MHz}$ for solid and $100 \mathrm{kHz}$ for dotted lines. }
    \label{fig:SigThroughTrans}
\end{figure}

In Fig.~\ref{fig:SigThroughTrans} we show the signal flux spectral density for three different DM masses, fixing $g_{\rm eff} = 10^{-10}$ and showing three choices for the detector bandwidth. This demonstrates how the detector response affects the signal as a function of frequency and bandwidth.

\end{document}